\documentclass[english]{article}
\usepackage[latin9]{inputenc}
\usepackage{amsmath}
\usepackage{multirow}
\usepackage{placeins}
\usepackage{babel}



\usepackage{babel}
\begin{document}

\title{Study of the Q.Q Interaction- Single Particle Behaviour to Elliott's
Rotations - J(J+1), J(J-1) and J(J+3) Spectra}

\author{Arun Kingan, Xiaofei Yu and Larry Zamick \\
 Department of Physics and Astronomy, \\
 Rutgers University, Piscataway, New Jersey 08854}
\maketitle
\begin{abstract}
We perform shell model calculations using a quadrupole-quadrupole
interaction (Q.Q).We show results in single j shell spaces and the
full S-D shell. We show that one gets useful results with Q.Q in both
spaces. We emphasize the importance of the choice of single particle
energies in order to obtain the results of Elliott using a Q.Q interaction
without the momentum terms. We show a J(J+1) spectrum for a ground
state band but with B(E2)'s different from the rotational model. We
also show results not found in textbooks such as J (J+1), J(J-1) and
J(J+3) excited bands. We find spectra starting with J=0 which have
both even J and odd J members. 
\end{abstract}

\section{Introduction}

Our goal is to systematically reexamine the Elliott SU3 model{[}1,2,3,4{]}
with a few variations. We use a Q.Q interaction but without the momentum
terms. When using the shell model to reproduce Elliott one has to
introduce a specific single particle splitting, some of which includes
interactions of the valence particle with the core {[}4,5{]}. But
before doing all this we show that the Q.Q interaction is very useful
in single j shell calculations and show an example where agreement
with experiment is remarkable. Back to Elliott we look not only at
spectra but systematics of quadrupole moments and B(E2)'s. A difference
from the Elliott et al. papers above{[}1,2,3,4{]} is that they mainly
emphasize the orbital parts of the wave functions--so spin or isospin
quantum numbers are not shown.Examples are shown only for S=0 states
i.e. J=L. In our shell model approach we get the full package-states
with all possible spins and all possible isospins.

\section{The Q.Q interaction in the single j shell.}

The interaction we use is -$\chi$ Q.Q = -$\chi$ $\sqrt{5}$ {[}(r$^{2}$
Y$^{2}$)$_{i}$(r$^{2}$ Y$^{2}$)$_{j}${]}$^{0}$ In evaluating
energies, unless specified otherwise, we set $\chi$ b$^{4}$ to 1
MeV. Alternately one can say that the energy is in units of $\chi$b$^{4}$.

In the single j shell one definitely does not get a rotational spectrum.
In Table I we compare the spectrum of even J states in $^{52}$Fe
resulting from using Q.Q in a single j shell (f$_{7/2}$), (2 proton
holes and 2 neutron holes in the f$_{7/2}$ shell). The strength of
the interaction was adjusted so the energy of the J=2$^{+}$ state
agreed with experiment. Given the simplicity of the interaction and
the smallness of the model space the agreement is remarkable. Note
that the energy of the J=12$^{+}$ state is lower than J=10$^{+}$
and this is reproduced in the calculation. Thus the J=12$^{+}$is
a very long lived isomeric state. Many other spin gap isomers are
predicted using Q.Q in a single j shell. For$^{52}$ Fe and $^{53}$
Co the J=15/2 $^{+}$,17/2$^{+}$ and 19/2$^{+}$ states are at 4.448
,5.145 and 3.588 MeV respectively. Thus the J=19/2$^{+}$ state is
isomeric. In $^{96}$ Cd the J=14$^{+}$,15 $^{+}$ and 16$^{+}$
states are at 4.138,4.201 and 3.483 MeV. Thus the J=16$^{+}$ state
is predicted to be isomeric with Q.Q in single j. In $^{96}$Ag the
J=10$^{+}$, 11$^{+}$and 12$^{+}$ states are predicted to be at
2.535, 3.089 and 2.482 MeV respectively. Thus the J=12$^{+}$ state
is predicted to be isomeric. These isomerisms have been verified experimentally.

\begin{table}[h]
\global\long\def\thetable{I}
 \centering \caption{Single j shell spectrum of $^{52}$Fe--Q.Q vs. experiment}
\begin{tabular}{c c c}
\hline 
J  & Q.Q  & EXP \\
\hline 
0  & 0  & 0  \\
2  & 0.849  & 0.849  \\
4  & 2.094  & 2.384  \\
6  & 3.982  & 4.325  \\
8  & 5.996  & 6.360  \\
10  & 7.389  & 7.382  \\
12  & 7.168  & 6.958  \\
\hline 
\end{tabular}
\end{table}

Admittedly we have chosen the best example. In $^{44}$Ti (2 protons
and 2 neutrons in the f$_{7/2}$ shell) the J=12$^{+}$ state is slightly
above the J=10$^{+}$state although it is still isomeric. Still, using
the Q.Q in a wide variety of single j shell calculations gives a reasonable
good start and is useful for orientation in regions where there is
not enough data to get the 2 body matrix elements from experiment.
For example in the g$_{9/2}$ shell the spectrum of the 2-hole system
$^{98}$In is not known.

For completeness we briefly mention some previous results involving
the Q.Q interaction which so far are not well understood. In ref {[}6{]}
it was noted that for identical particles in the g$_{9/2}$ shell
seniority is in general not a good quantum number. However, it is
for a limited set of interactions which do conserve seniority such
as the delta interaction. If we compare the spectra of 3 neutrons
in the g$_{9/2}$ shell with that of 5 neutrons we find E(21/2)-E(3/2)
is the same in the 2 cases. Now the Q.Q interaction does not conserve
seniority for identical particles in the g$_{9/2}$ shell or beyond.
What is of interest here is that the above splitting is equal in magnitude
but opposite in sign for 3 and 5 particles.

Another unproven result by the same authors {[}8{]} for a system of
2 protons and 2 neutrons in a given shell - one finds when using a
Q.Q interaction that some (but not all) T=2 states are degenerate
in energy with some T=0 states. Of particular interest in the g$_{9/2}$
shell is the degeneracy of a unique J=4 T=2 state with seniority v=4
with a J=4 T=0 state also with seniority v=4. The J=4 T=2 v=4 state
appears no matter what interaction is used even though in general
seniority is not conserved for identical particles in the g$_{9/2}$
shell. Even more surprising is that with a Q.Q interaction there is
a T=0 J=4 state with a definite seniority and it is degenerate with
the unique J=4 T=2 state. In general, with any interaction, not just
Q.Q, when one has mixed protons and neutrons seniority is not conserved
in any shell.

For completeness we note that Zamick and Harper {[}9{]} showed that
for 2 protons and 2 neutrons in a single j shell there is a very high
overlap between the wave functions arising from a Q.Q interaction
and properly symmetrized unitary 9j coefficients (U9j).

Another interesting feature of Q.Q in a single j shell calculation
is that the spectrum of a particle-hole is ``upside down'' the spectrum
of 2 particles. This is actually true for any multipole-multipole
interaction, as noted by Talmi{[}10{]}.The empirical orders of the
spins of energy levels, from low to high, for $^{42}$Sc ( 2 particles)
is 0,1,7,3,5,2,4,6. For $^{48}$Sc (proton particle-neutron hole)
the ordering is 6,5,4,3,7,1,2,0.

\section{Elliott Model- Single Particle Energies and Degeneracies }

In contrast to the previous section we here consider the use of Q.Q
to produce rotational states in the shell model. We refer of course
to the Elliott SU(3) model {[}1,2,3,4{]}.We now have to consider all
configurations in a major shell. Although this model has been well
studied we wish to emphasize certain aspects which are perhaps not
so familiar, especially the choice of single particle energies in
a formulation where we do not include the momentum terms in the interaction.
We use the simple Q.Q interaction described above. The numbers are
expressed in units of $\chi$ b$^{4}$ where b is the oscillation
length parameter (or if you like we set the value of $\chi$ b$^{4}$to
one).

The Elliott formula for the energies is

\begin{equation}
E(SU(3)) = \chi' [-4(\lambda^{2}+\mu^{2}+\lambda \mu+3(\lambda+\mu))]+3\chi' L(L+1)
\end{equation}

\noindent where $\chi$'= 5b$^{4}$/(32$\pi$) $\chi$.

To get Elliott's SU(3) results in the shell model one has to introduce
a single particle energy splitting {[}4,5{]}

\begin{equation}
E(L_{2}) -E(L_{1})=3\chi'[L_{2}(L_{2}+1)-L_{1}(L_{1}+1)]
\end{equation}

The splitting is 18 $\chi$' in the S-D shell and 30 $\chi'$ in the P-F shell.
Note that the bigger L single particle level is at a higher energy
than the smaller, i.e. D is higher than S in the S-D shell and F is
at a higher energy than P in the P-F shell. This may go against experiment
but if one wants to get Elliott's results that is what one has to
do. As noted by Zamick et al. {[}5{]} and by Moya de Deguerra et al.
{[}6{]} when one uses the simple Q.Q interaction (without Elliott's
momentum terms) 2/3 of the splitting comes from the diagonal part
of the Q.Q interaction and 1/3 comes from the particle core interaction.
One can say that for the single nucleon configuration (e.g. $^{17}$O
or $^{41}$Ca), one also has a rotational band consisting of 2 states
L=0 and L=2 in the S-D shell and L=1 and L=3 in the P-F shell. See
also discussions of momentum term removals my Talmi {[}10{]}.

Before continuing we note that there have of course been many developments
since the works of Elliott including higher configuration admixtures
out of the S-D shell and works on higher shells e.g. P-F. Some selected
works are refs{[}11-17{]}. Our intent here is quite different. We
want to take a hard look at the Elliott model in it's simplest form
and see if there are some interesting features worth pointing out.
The heading of the next section indicates that there are.

We will call this work SMQ.Q (SM=Shell Model), so as to make the distinction
of working without and with momentum terms.

\section{J (J+1), J(J-1) and J(J+3) Spectra}

In Table II we show contrasting spectra. The familiar ground state
band has a J(J+1) spectrum with only even J's. But the lowest excited
bandhead at 5.073 MeV is multidegenerate. For J=1 there are 3 states
at this energy, one with isospin T=0 and two with T=1. Here we show
three bands that can be extracted. We will here consider only T=0
bands.

\begin{table}[h]
\global\long\def\thetable{II}
 \centering \caption{Ground state and excited state band energies in units of $\chi$ b$^{4}$}
\begin{tabular}{c c c c c}
\hline 
J  & Ground Band  & Excited Band 1  & Excited Band 2  & Excited Band 3 \\
\hline 
0  & 0  &  &  & 5.073 \\
1  &  & 5.073  &  & 5.670\\
2  & 0.895  & 5.670  & 5.073  & 6.565 \\
3  &  & 6.565  & 5.607  & 7.751 \\
4  & 2.984  & 7.759  & 6.565  & 9.251 \\
5  &  & 9.251  & 7.759  & 11.041 \\
6  & 6.266  & 11.041  & 9.251  & 13.340 \\
7  &  & 13.130  & 11.041  & \\
8  & 10.743  &  & 13.130  & \\
\hline 
\end{tabular}
\end{table}

The ground band energies are given by E(J) = 0.149 J(J+1). The energies for excited bands 1, 2, and 3 are given respectively by
\begin{equation}
E(J)= 4.772 + 0.149 J(J+1)
\end{equation}
\begin{equation}
E(J)= 4.772 + 0.149 J(J-1)
\end{equation}
\begin{equation}
E(J)= 5.073 + 0.149 J(J+3)
\end{equation}
\noindent Notice that for all bands the coefficient of $J^{2}$ is the same,
namely 0.149. This means that these bands all have the same moments
of inertia. The $\lambda$$\mu$ values for the ground band are (8,0)
and for the 3 excited bands (6,1).

One cannot help but notice that excited band 2 looks the same as excited
band 1 except that J is shifted up by one unit. Likewise, band 3 has
J shifted down by one unit relative to excited band 1. This suggests
reorientations of L and S for these 3 bands. An unusual feature of
band 3 is that it starts with J=0 but, unlike the ground state band
it includes both even and odd J's.

The pattern im Table 2 suggests that the {[} L S{]}J configurations
of the 3 excired bands are respectively {[}L 1{]}J=L , {[}L 1 {]}J=L+1
and {[}L 1{]}J=L-1. We will examine electric quadrupole and magnetic
dipole properties to verify these assignments.

\section{B(E2)'s and Q(2$^{+}$) in the Elliott Model.}

In this work we take the effective charges to be 1.5 for the proton
and 0.5 for the neutron. In Table III we list the B(E2)'s along the
ground state band in the full space for nuclei in the S-D shell. They
are in units of e$^{2}$fm$^{4}$. We also show the same results in
a reduced space (Table IV) where only s$_{1/2}$ and d$_{5/2}$ subshells
are allowed (no d$_{3/2}$). This gives us a sense of how increasing
configurations affects collectivity. The B(E2)'s in the full space
(Table III)) are substantially larger than in the reduced space. Whereas
we get a perfect J(J+1) spectrum in Table III, we get a more compressed
spectrum in Table IV and the J(J+1) fit is only approximate. Not surprisingly,
the static quadrupole moments for J=2 and 4 are larger in magnitude
in the full space than those in the reduced space. In Vol 2 of their
book Bohr and Mottelson give analytic formulas for Q and B(E2) {[}18{]}.

\[
\langle K=0,I_{2}||\mu(2)||K=0,I_{1}\rangle
\]
\begin{equation}
=\left(\frac{5}{16\pi}\right)^{1/2}\frac{\hbar}{M\omega_{o}}(2\lambda+3)(2I_{1}+1)^{1/2}\langle I_{1}020|I_{2}0\rangle\times\{\begin{cases}
1 & I_{2}=I_{1}\\
\left(1-\left(\frac{2I_{1}+3}{2\lambda+3}\right)^{2}\right)^{1/2} & I_{2}=I_{1}+2
\end{cases}
\end{equation}

We acknowledge early work on B(E2)'s with the Elliott model by Strottmann
{[}19{]}.

\begin{table}[h]
\global\long\def\thetable{III}
 \centering \caption{Quadrupole moments (e fm$^{2}$ ) and B(E2)'s (e$^{2}$ fm$^{4}$)
for the ground state band in the Elliott model-full S-D Space.}
\begin{tabular}{c c c c}
\hline 
Energy  & J  & Q(J)  & B(E2)J$\rightarrow J$+2 \\
\hline 
0  & 0  & 0  & 427 \\
0.8952  & 2  & -18.96  & 194 \\
2.9841  & 4  & -24.13  & 132 \\
6.2665  & 6  & -26.55  & 91 \\
10.7426  & 8  & -27.95  &  \\
\hline 
\end{tabular}

\global\long\def\thetable{IV}
 \centering \caption{Same as Table III but in reduced space - only s$_{1/2}$ and d$_{5/2}$
subshells included (no d$_{3/2}$).}
\begin{tabular}{c c c c}
\hline 
Energy  & J  & Q(J)  & B(E2)J$\rightarrow J$+2 \\
\hline 
0  & 0  & 0  & 290 \\
0.9983  & 2  & -14.75  & 121 \\
2.9962  & 4  & -15.34  & 79\\
5.6076  & 6  & -13.46  & 44 \\
8.091  & 8  & -11.81  &  \\
\hline 
\end{tabular}
\end{table}

We note that the B(E2) from the lowest 2$^{+}$ state to the J=0 ground
state is strong with the Q.Q interaction. The results are not dissimilar
to what one obtains with realistic interactions. We make a comparison
with the rotational model of Bohr and Mottelson{[}17{]} for which
the following formulas hold:

\begin{equation}
B(E2,K J_{2}\rightarrow K J_{1}) = 5/(16\pi) e^{2}Q_{0}^{2}\textless{}
J_{1}2 K0\textbar{} J_{2}K\textgreater{}^{2}
\end{equation}

\begin{equation}
Q(J)=(3 K^{2}-J(J+1))/((J+1)(2J+3)) Q_{0}
\end{equation}

For J=2:

Q(K=0)=-2/7 Q$_{0}$

Q(K=2) = +2/7 Q$_{0}$ They are equal and opposite.

For J=0 K=0 $\rightarrow$J=2 K=0 we have 

\begin{equation}
-1.1039 Q(J=2)/ \sqrt{B(E2)} = 1.013. 
\end{equation}

In the rotational model it would be one. How does the Elliott
model compare with the rotational model?

In his second paper {[}2{]} Elliott says that the quadrupole moments
in a \char`\"{} K=0 rotational band\char`\"{} are identical to those
of the rotational model but the B(E2)'s are not. In Table III we confirm
this for the ground state band of $^{20}$Ne.

In Table V we show selected quadrupole moments of 2+ states and B(E2)'s
from the 0$_{1}$$^{+}$ ground state to several 2+ states.

\begin{table}[h]
\global\long\def\thetable{V}
 \centering \caption{E(2$_{n}$$^{+}$)MeV, Q(2$_{n}$) e fm$^{^{2}}$ and B(E2)0$_{1}$$\rightarrow$2$_{n}$
e$^{2}$fm$^{4}$in $^{20}$Ne}
\begin{tabular}{c c c c}
\hline 
E (2$^{+}$)  & T  & Q(2$_{n}$)  & B(E2)0$_{1}$$\rightarrow$2$_{n}$ \\
\hline 
0.895  & 0  & -18.96  & 427.0\\
5.073  & 0  & 8.05  & 0 \\
5.670  & 0  & -5.60  & 0 \\
6.565  & 0  & -7.38  & 0 \\
6.565  & 1  & -8.43  & 12.56 \\
8.356  & 0  & 0  & 0 \\
\hline 
\end{tabular}
\end{table}

Note that although the strongest B(E2) is the intraband transition
from 0$_{1}$ to 2$_{1}$(427 e$^{2}$fm$^{4}$), there is a finite,
albeit weak, B(E2) to a T=1 state at 6.565 MeV. It should be pointed
out that if we had chosen effective charges that were the same for
the proton and the neutron, i.e. the isoscalar choice, there would
not be any B(E2) strength to states at 6.565 MeV or for that matter
to any states except the one at 0.895 MeV. Hence only 2 non-zero B(E2)'s,
one to a T=0 one to a T=1 final state.

There are no other finite B(E2)'s from the 0$_{1}$ ground state,
even to 2$^{+}$ states not shown. If we look at transitions from
the 2$_{1}$$^{+}$ state to 0$_{n}$$^{+}$ states there is only
a single non-zero transition 2$_{1}$$^{+}$ to 0$_{1}$$^{+}$ (427.0/5=85.4
e$^{2}$fm$^{4}$).The B(E2)'s to all other 0$^{+}$ states vanish.
This is true even if the effective charges of the neutrons and protons
are different..

As shown in Table V the quadrupole moment of the 2$_{1}$$^{+}$ state
is negative, consistent with a prolate deformation for a K=0 band.
There is a change of sign at 5.073 MeV consistent with a K=2 prolate
band. Since there are 3 degenerate states at 5.670 MeV there is arbitrary
as to how we distribute Q and B(E2) between the 2 T=0 degenerate states
Clebsch-Gordan coefficients.

\begin{table}[h]
\global\long\def\thetable{VI}
 \centering \caption{B(E2)'s in excited band 1}
\begin{tabular}{c c c | c c c}
\hline 
\multicolumn{6}{c}{Band 1}\\
\hline 
\hline 
\multicolumn{3}{c|}{$\Delta$ J = 2} & \multicolumn{3}{c}{$\Delta$ J = 1}\\
\hline 
$J_{i}$ {[}$E_{i}${]}  & $J_{f}$ {[}$E_{f}${]}  & B(E2)  & $J_{i}$ {[}$E_{i}${]}  & $J_{f}$ {[}$E_{f}${]}  & B(E2) \\
\hline 
1 {[}5.07{]}  & 3 {[}5.67{]}  & 53.11  & 1 {[}5.07{]}  & 2 {[}5.67{]}  & 10.99 \\
 & 3 {[}6.60{]}  & 86.13  &  & 2 {[}6.56{]}  & 43.24 \\
2 {[}5.07{]}  & 4 {[}6.56{]}  & 18.09  & 2 {[}5.67{]}  & 3 {[}6.56{]}  & 41.85 \\
 & 4 {[}7.79{]}  & 76.4  &  & 3 {[}7.82{]}  & 15.17 \\
3 {[}6.56{]}  & 5 {[}9.25{]}  & 97.39  &  & 3 {[}9.25{]}  & 1.3 \\
 & 5 {[}7.75{]}  & 2.13  & 3 {[}6.56{]}  & 4 {[}7.75{]}  & 11.65 \\
4 {[}7.70{]}  & 6 {[}9.25{]}  & 2.78  &  & 4 {[}9.25{]}  & 10.85 \\
 & 6 {[}11.04{]}  & 48.98  & 4 {[}7.79{]}  & 5 {[}9.25{]}  & 2.78 \\
6 {[}11.04{]}  & 8 {[}13.13{]}  & 0.82  &  & 5 {[}11.08{]}  & 49.98 \\
 &  &  & 5 {[}9.25{]}  & 6 {[}11.08{]}  & 3.04 \\
 &  &  & 6 {[}11.08{]}  & 7 {[}13.13{]}  & 17.19 \\
\hline 
\end{tabular}
\end{table}

\begin{table}
\global\long\def\thetable{VII}
 \centering \caption{B(E2)'s in excited band 2 e$^{2}$fm$^{4}$}
\begin{tabular}{c c c | c c c}
\hline 
\multicolumn{6}{c}{Band 2}\\
\hline 
\hline 
\multicolumn{3}{c|}{$\Delta$ J = 2} & \multicolumn{3}{c}{$\Delta$ J = 1}\\
\hline 
$J_{i}$ {[}$E_{i}${]}  & $J_{f}$ {[}$E_{f}${]}  & B(E2)  & $J_{i}$ {[}$E_{i}${]}  & $J_{f}$ {[}$E_{f}${]}  & B(E2) \\
\hline 
2 {[}5.07{]}  & 4 {[}6.56{]}  & 97.22  & 2 {[}5.07{]}  & 3 {[}5.67{]}  & 66.13 \\
3 {[}5.67{]}  & 5 {[}7.75{]}  & 79.83  &  & 3 {[}6.56{]}  & 25.3 \\
4 {[}6.56{]}  & 6 {[}9.25{]}  & 100  & 3 {[}5.67{]}  & 4 {[}6.56{]}  & 66.05 \\
5 {[}7.75{]}  & 7 {[}11.04{]}  & 50.49  &  & 4 {[}7.79{]}  & 10.89 \\
6 {[}9.25{]}  & 8 {[}13.13{]}  & 58.6  & 4 {[}6.56{]}  & 5 {[}7.75{]}  & 14.96 \\
 &  &  &  & 5 {[}9.30{]}  & 8.46 \\
 &  &  & 5 {[}7.75{]}  & 6 {[}9.25{]}  & 30.55 \\
 &  &  &  & 6 {[}11.04{]}  & 2.92 \\
 &  &  & 6 {[}9.25{]}  & 7 {[}11.04{]}  & 3.35 \\
 &  &  &  & 7 {[}13.13{]}  & 2.47 \\
 &  &  & 7 {[}11.04{]}  & 8 {[}13.13{]}  & 18.27 \\
 &  &  & 7{*} {[}12.76{]}  & 8{*} {[}14.57{]}  & 28.28 \\
\hline 
\end{tabular}

\global\long\def\thetable{VIII}
 \centering \caption{B(E2)'s in excited band 3}
\begin{tabular}{c c c | c c c}
\hline 
\multicolumn{6}{c}{Band 3}\\
\hline
\hline
\multicolumn{3}{c|}{$\Delta$ J = 2} & \multicolumn{3}{c}{$\Delta$ J = 1}\\
\hline
$J_{i}$ \{{[}\}$E_{i}$\{{]}\}  & $J_{f}$\{{[}\}$E_{f}$\{{]}\}  & B(E2)  & $J_{i}$\{{[}\}$E_{i}$\{{]}\}  & $J_{f}$\{{[}\}$E_{f}$\{{]}\}  & B(E2) \\
\hline
0 {[}5.07{]}  & 2 {[}5.67{]}  & 116.7  & 1 {[}5.67{]}  & 2 {[}6.56{]}  & 55.28 \\
 & 2 {[}6.56{]}  & 130.5  & 2 {[}6.56{]}  & 3 {[}7.75{]}  & 13.30 \\
1 {[}5.67{]}  & 3 {[}6.56{]}  & 28.68  & 3 {[}7.75{]}  & 4 {[}9.25{]}  & 29.88 \\
 & 3 {[}7.75{]}  & 91.45  & 4 {[}9.25{]}  & 5 {[}11.04{]}  & 3.22 \\
2 {[}6.56{]}  & 4 {[}9.25{]}  & 108.4  & 5 {[}11.04{]}  & 6 {[}13.34{]}  & 18.11 \\
3 {[}7.75{]}  & 5 {[}11.04{]}  & 53.25  &  &  & \\
4 {[}9.25{]}  & 6 {[}13.34{]}  & 61.13  &  &  & \\
\hline 
\end{tabular}
\end{table}

\FloatBarrier

In Tables VI, VII, and VIII we give results for B(E2)'s for the excited
bands 1, 2, and 3. We show results only to final states of isospin
Tf=0 but it should be noted that there are finite, albeit small, B(E2)'s
to Tf=1 states. We use these as starting energies and angular momenta
the ones shown in Table II-those of a J(J-1) spectrum. The results
for $\Delta$J=2 in band 2 show a simple behavior with strong B(E2)'s
to higher J states and with concentration of the strength to one final
states. However for delta J=1 the behavior is more complex. One gets
ranches to mainly 2 final states of the same angular momentum e.g.
3{[}5.67{]} to 4{[}6.56{]} and 2 4{[}7.79{]} with respective strengths
of 66.05 and 10.89 e$^{2}$fm$^{4}$. For excited band 1 we get fragmentation
for both delta J=2 and delta J=1

We can reduce the expression for the B(E2) from configuration {[}L$_{i}$1{]}J$_{i}$
to {[}L$_{f}$ 1{]} J$_{f}$

\begin{equation}
B(E2) = (2J_{f}+1)/((2J_{i}+1)(2L_{f}+1)){[}U(2 fL_{i}J_{f} 1; J_{i}L_{f})*\textless{}L_{f}\textbar{}\textbar{}E^{2}\textbar{}\textbar{}L_{i}\textgreater{}{]}^{2}
\end{equation}

Here U is the unitary Racah coefficient. One can use this to get ratios
of B(E2)'s of transitions which have the same L$_{i}$ and L$_{l}$.
In those cases the reduced matrix elements drop out. For example in
transtions from 5.67 MeV to 6.56 MeV, the value from J=3 to J=4 (Band
1) is 66.05 and from J=2 to J=3 (Band 2) is 41.85 so the ratio is
1.54. One can easily verify that one gets the same ratio from the
above expression.

.

\begin{table}[h]
\global\long\def\thetable{IX}
 \centering

\caption{Ratios in the ground state band of $^{20}$Ne-- Q(J$_{f}$)/Q(2) and
B(E2) J$_{f}$-2$\rightarrow$J$_{f}$/B(E2)0$\rightarrow2$}
\begin{tabular}{c | c c c c}
\hline 
J$_{f}$  & 2  & 4  & 6  & 8\\
\hline 
Q(J$_{f}$)/Q(2)Elliott  & 1  & 1.273  & 1.4  & 1.474\\
Q(J$_{f}$)/Q(2)Rotational  & 1  & 1.273  & 1.4  & 1.474\\
B(E2)J$_{f}$/B(E2)$_{2}$ Elliott  & 1  & 0.456  & 0.310  & 0.167\\
B(E2)J$_{f}$/B(E2)$_{2}$ Rotational  & 1  & 0.515  & 0.455  & 0.430\\
\hline 
\end{tabular}

\global\long\def\thetable{X}
 \centering \caption{Comparison of quadrupole moments (e fm$^{2}$) in the SMQ.Q model
and the rotational model for the lowest ''K=1'' band.}
\begin{tabular}{c c c c}
\hline 
Energy MeV  & J  & Q(J) SMQ.Q  & Q(J) rotational K=1 \\
\hline 
\hline 
5.073  & 1  & -3.70  & 0.49 \\
5.668  & 2  & -5.60  & -7.00 \\
6.561  & 3  & -8.04  & -12.25 \\
7.751  & 4  & -17.90  & -15.15 \\
9.251  & 5  & -13.84  & -18.20 \\
11.041  & 6  & -23.14  & -18.96 \\
13.130  & 7  & -16.55  & -19.77 \\
\hline 
\end{tabular}
\end{table}

\begin{table}
\global\long\def\thetable{XI}
 \centering \caption{Comparison of quadrupole moments (e fm$^{2}$) in the SMQ.Q model
and the rotational model for the lowest ''K=2'' band}
\begin{tabular}{c c c c}
\hline 
Energy MeV  & J  & Q(J) SMQ.Q  & Q(J) rotational K=2 \\
\hline 
\hline 
5.073  & 2  & 8.02  & 16.65 \\
5.668  & 3  & -10.77  & 0 \\
6.561  & 4  & -10.93  & -8.43 \\
7.751  & 5  & -21.02  & -13.46 \\
9.251  & 6  & -15.20  & -16.66 \\
11.041  & 7  & -24.91  & -18.86 \\
13.130  & 8  & -17.46  & -20.46 \\
\hline 
\end{tabular}

\global\long\def\thetable{XII}
 \centering \caption{Comparison of quadrupole moments (e fm$^{2}$) in the SMQ.Q model
and the rotational model for the lowest ''K=0'' band}
\begin{tabular}{c c c c}
\hline 
Energy MeV  & J  & Q(J) SMQ.Q  & Q(J) rotational K=2 \\
\hline 
\hline 
5.073  & 0  & 0  & 0 \\
5.668  & 1  & -3.982  & -3.982 \\
6.561  & 2  & -7.383  & -5.689 \\
7.751  & 3  & -17.29  & -6.637 \\
9.251  & 4  & -13.57  & -7.240 \\
11.041  & 5  & -22.90  & -7.658 \\
13.130  & 6  & -16.39  & -7.964 \\
\hline 
\end{tabular}
\end{table}

\FloatBarrier

We confirm in Table IX the statement by Elliott{[}2{]} that the quadrupole
moments in his model are identical to those of the rotational model
for the ground band, which in the rotational model has K=0.We also
confirm his statement that the B(E2)'s for the ground state are different.
In fact, they are quite different. Elliott's B(E2)'s drop off much
faster with J than those of the rotational model. The same thing happens
in shell model calculations with more realistic interactions {[}20{]}.

In Table X we compare the quadrupole moments for excited band 1, as
calculated in the SMQ.Q model with those of a K=1roational band. We
show results for T=0 bands in Tables X through XII because there are
less degeneracies for T=0 than for T=1. A least squares fit was made
to minimize the deviations of the 2 models. In Table XI a similar
comparison was made for excited and 2 with a K=2 rotational band.
There are many differences. In the rotational model there is a monotonic
decrease in the quadrupole moments with J but this is not the case
in the SMQ.Q model. There are however some similarities such as the
change of sign in Table XI as one goes from J=2 to higher J. The fact
that the J=1 state of excited band 1 has a quadrupole moment of opposite
sign to that of the J=2 state of band 2 favors an L=1 assignment to
these to states rather than L=2.

\section{Magnetic moments as identifiers of configurations in LS coupling}

In this work much of the pre-Elliott work by E. P. Wigner comes into
play {[}21{]}.We show results in Table XIII for magnetic moments of
levels in excited bands 1, 2, and 3 using NuShellX. We can easily
associate these with LS coupling wave functions {[}LS{]}J by using
the following expression applicable to T=0 states:

\begin{equation}
\mu=(G_{l}[J(J+1)+L(L+1)-S(S+1)]+G_{s}[J(J+1)-L(L+1)+S(S+1)])/(2*(J+1))
\end{equation}

\begin{table}[h]
\global\long\def\thetable{XI}
 \centering \caption{Magnetic Moments for Bands 1, 2, 3 using Nushellx}
\begin{tabular}{c c c c c c c}
\hline 
E  & \multicolumn{2}{l}{Band 1} & \multicolumn{2}{l}{Band 2} & \multicolumn{2}{l}{Band 3}\\
\hline 
 & {[}LS{]}J=L  & $\mu$  & {[}LS{]}J=L+1  & $\mu$  & {[}LS{]}J=L-1  & $\mu$ \\
\hline 
\hline 
5.07  & {[}11{]}1  & 0.69  & {[}11{]}2  & 1.38  & {[}11{]}0  & \\
5.70  & {[}21{]}2  & 1.126  & {[}21{]}3  & 1.88  & {[}21{]}1  & 0.310 \\
6.56  & {[}31{]}3  & 1.524  & {[}31{]}4  & 2.38  & {[}31{]}2  & 0.747 \\
7.75  & {[}41{]}4  & 2.076  & {[}41{]}5  & 2.88  & {[}41{]}3  & 1.215 \\
9.26  & {[}51{]}5  & 2.563  & {[}51{]}6  & 3.38  & {[}51{]}4  & 1.696 \\
11.04  & {[}61{]}6  & 3.547  & {[}61{]}7  & 3.88  & {[}61{]}5  & 2.183 \\
13.13  & {[}71{]}7  & 4.042  & {[}71{]}8  & 4.28  & {[}71{]}6  & 2.675 \\
\hline 
\end{tabular}
\end{table}

For T=0 states the bare coupling values are. $G_{l}$=(1+0)/2 =0.5
$G_{s}$=(5.586-3.826)/2=0.88 For the ground state band S is equal
to zero, ({[}L 0{]} L) so the expression is simply $\mu$ =$G_{l}$
L (with J= L). For the excited bands the above formula for $\mu$
agrees with the NuShellX results when we attribute to band 1 the configuration
{[}LS{]} J=L and to band 2 {[}LS{]}J=L+1. Note that band 2 is a stretched
band so the formulas for $\mu$ is especially simple Mu=$G_{l}$ L+$G_{s}$.S=0.5
(J-1)+0.88.

The assigned configurations help to sharpen what was said in previous
sections. The fact that band 2 has a J(J-1) spectrum is due to the
spin independence of the interaction that is here used, as well as
the fact that $d_{3/2}$ and $d_{5/2}$ are degenerate. It costs no
energy to take the spin orientation in band 1 and stretch it out to
form band 2. Note that the results for the magnetic moments do not
depend on the SU(3) quantum numbers-only on L, S, and J.

Some of the patterns of the B(E2)'s can also be explained. For the
cases where $J_{f}$=$J_{i}$+2 there is a large fragmentation of
B(E2) strength in Band 1 but not in Band 2. In band 1 one can go from
{[}LS{]}L to {[}(L+1)S{]}L+2 and {[}(L+2) S{]}L+2 but from band 2
one can only go from {[}LS{]}L+1 to {[}L+2 S{]}L+3.

\section{CLOSING REMARKS}

We start with a technical point. The high degeneracies resulting from
the SMQ.Q (Elliott) model can lead to problems. One of the main ones
is that the isospin assignments for degenerate states can get mixed
up. We addressed this by raising the d$_{3/2}$ state 0.1 MeV above
d$_{5/2}$. This removed most degeneracies so we could distinguish
which states had isospins T=0. It also helps us keep a track of a
band when for higher spin new bands pop up.

We have here considered various aspects of the Q.Q interaction. It
serves as a reasonable interaction in small spaces where for example
it has the feature of level inversion at high spins which is seen
in many nuclei-e.g. $^{52}$Fe and other nuclei mentioned above. There
are also early discussions of other spin gaps by Auerbach and Talmi
{[}22,23{]}.

In a full space i.e. Elliott's SU(3) model {[}1,2,3{]}
we feel the simplest aspects of the model deserve further attention.
This model has sometimes been described as giving us rotations in
the shell model. While this is true of the spectrum one has to qualify
this statement when considering B(E2)'s and quadrupole moments, as noted by Elliott [1,2]. As
one goes to high spins the B(E2)'s in the rotational model seem to
flatten out but in the Elliott model, as indeed in the shell model
after a certain J they fall off. In the rotational model the quadrupole
moments are monotonically decreasing (i.e. becoming more negative)
but in the Elliott model this is not the case. Perhaps the best thing
to say is that the Elliott model gives us rotational behavior with
shell model modifications.

 As mentioned in the intorduction the early
papers of Elliott and Harvey {[}1,2,3,4{]} emphasize the orbital parts
of the wave functions and although LS coupling is mentioned one does
not see spin or isospin labels. In our shell model approach, SMQ.Q
, using NUshellX {[}24 {]} we necessarity get complete wave functions
-orbital and spin combined , and the isospin quantum numbers as well.
And intersting results come when we look at the behavior as a function
of J rather than L.

By looking hard at this model, resisting the temptation to modify
single particle energies to fit experiment, we uncover interesting
new features of this such as the J(J-1) and J(J+3) spectra. We showed
this for bands for which the lowest L value was one, but we get the
same struture for any higher L. Perhaps the most interesting result
is that we get spectra starting from J=0 which have both even J and
odd J members i.e. J=0,1,2,3,4,5,6 for the configuration {[}L S=1{]}J=L-1.
Although we have not explicitly tried to fit experimental data we
have here suggestions for experiment, namely to look for these different
patterns that we have found, or at lease for remnants of these patterns.
We hope both experimentalists and theorists will continue in this
fascinating pursuit.

\section{ACKNOWLEDGEMENTS}

One of us (L.Z.) did some of this work during a visit at the Weizmann
Institute. He thanks Micahel Kirson , Igal Talmi and Naftali Auerbach
for their hospitality and input. One of us (A.K.) was supporedt by
a Richard J. Plano award for the summer of 2017. X.Yu is now in the
graduate program at the University of Chicago. We than Alex Brown
and Shadow Robinson for valuable comments.

\end{document}